\documentstyle[12pt]{article}
\newcommand{\mrm}[1]{\mbox{\rm #1}}
\newcommand{\half}{{1\over 2}}
\newcommand{\bla}{\hspace{1cm}}
\newcommand{\eq}[1]{eq.~(\ref{#1})}
\newcommand{\rfn}[1]{(\ref{#1})}

\newcommand{\abs}[1]{\left| #1\right|}
\renewcommand{\Re}[1]{\mathop{\mrm{Re}}\left\{ #1 \right\}}
\newcommand{\ltap}[1]{\ \raisebox{-.4ex}{\rlap{$\sim$}} \raisebox{.4ex}{$<$}\ }
\newcommand{\gtap}[1]{\ \raisebox{-.4ex}{\rlap{$\sim$}} \raisebox{.4ex}{$>$}\ }

\renewcommand{\titlepage}{\clearpage%
\setcounter{footnote}{0}%
\thispagestyle{empty}\pagestyle{plain}\pagenumbering{arabic}%
\kern1mm
\vskip15mm\normalsize}
\newcommand{\docnum}[1]{\hbox to \hsize{\hskip123mm\hbox{#1}\hss}}
\renewcommand{\date}[1]{\hbox to \hsize{\hskip123mm\hbox{#1}\hss}}

\renewcommand{\title}[1]{\vskip1em\begin{center}\Large\bf#1
\end{center}\vskip2.5em}
\renewcommand{\author}[1]{\vskip0.5em{\bf #1}\vskip0.5em}
\newcommand{\inst}[1]{\vskip0.3em{ #1}\vskip0.5em}
\renewcommand{\abstract}{\begin{center}{\bf Abstract}
\end{center}\quotation}

\newcommand{\anotfoot}[2]{\vfill\noindent\underline{\hspace{6cm}}
\par\noindent #1) #2}
\newcommand{\anotfootnb}[2]{\par\noindent #1) #2}

\begin{document}
\begin{titlepage}
\docnum{CERN-TH.7230/94}
\vspace{0.5cm}
\title{Bounding Effective Operators at the One-Loop Level:  \\
The Case of Four-Fermion Neutrino Interactions}
\begin{center}
\author{Mikhail Bilenky$^{*)}$}
\inst{DESY-IfH, Platanenallee 6, D-15738 Zeuthen, Germany}
\inst{and}
\author{Arcadi Santamaria$^{**)}$}
\inst{TH Division, CERN, 1211 Gen\`eve 23, Switzerland}
\end{center}
\vspace{0.25cm}
\begin{abstract}
The contributions of non-standard four-neutrino contact interactions
to  electroweak observables are considered at the one-loop level by
using the effective quantum field theory.
The analysis is done
in terms of three unknown parameters:
the strength of the non-standard neutrino interactions, $\tilde{F}$,
an additional derivative coupling needed to renormalize
the divergent contributions that appear when the four-neutrino
interactions are used at the loop level and a non-standard
non-derivative $Z$-${\bar\nu} \nu$ coupling.
Then, the precise measurements of the invisible
width of the $Z$-boson at LEP and
the data on the neutrino deep-inelastic scattering yield the result
$\tilde{F} = (-100 \pm 140) G_F$.
Assuming that there are no
unnatural cancellations between the contributions of the three effective
couplings a much stronger bound is obtained: $\abs{\tilde{F}}\ltap~ 2 G_F$,
which is a factor 200 better than the one obtained in previous analyses
based on tree level calculations.
 \end{abstract}
 \vspace{0.25cm}
\vfill\noindent
CERN-TH.7230/94\\
May 1994
\anotfoot{*}{On leave of absence from
the Joint Institute for Nuclear Research, Dubna, Russia.}
\anotfootnb{**}{On leave of absence from Departament de
F\'{\i}sica Te\`orica,
Universitat de Val\`encia, and IFIC, Val\`encia, Spain.}
\end{titlepage}
\setcounter{page}{1}

\section{Introduction}
Until now the Standard Model \cite{SM} of electroweak interactions (SM)
has passed very successfully all precision experimental tests.
Especially intensive have been the studies of the
different four-fermion processes between leptons and quarks.
In the SM such reactions are mediated by the electroweak vector bosons.
The large body of experimental data collected at different energy scales
(from practically zero-energy up to the mass of the $Z$-boson) confirms that,
indeed, four-fermion interactions are mediated by gauge bosons.
\par
Nevertheless, there is a widespread belief among theoreticians that the current
theory of electroweak interactions is only an effective low-energy limit of
a more fundamental theory.
If there is some new dynamics beyond the SM,
it might result in some deviations from the SM predictions
for four-fermion processes.
For example, standard fermions could take part in  processes with the
exchange of some non-standard  intermediate state. If the mass
of the intermediate particle is larger than the Fermi scale,
then,  at the Fermi scale (and below) the new interaction can be described
by effective four-fermion operators suppressed by $1/M^2$, where $M$ is a
scale of the order of the mass of the heavy intermediate particle.
\par
Obviously one of the most elusive among the non-standard
four-fermion interactions is that which involves only neutrinos.
This type of interactions can naturally appear
in models with extra neutral gauge bosons or new scalars.
Here we will assume
simply that this interaction exists without asking about its particular
origin.

It is clear that if some ``secret'' neutrino interaction
(SNI) exists, it can only be tested indirectly.
\par
The first studies of possible non-standard $\nu$--$\nu$ interactions
were performed many years ago \cite{BIA64,BBP70}. In ref. \cite{BBP70}
different weak processes sensitive to such an interaction were investigated
for a SNI with pure vector form
\begin{equation}
{\cal L}^{\nu-\nu} =
 F (\bar{\nu}\gamma_\alpha \nu) (\bar{\nu}\gamma^\alpha \nu).
\label{effvect}
\end{equation}
In particular the SNI, which contributes to the decays
$\pi^+ \rightarrow e^+ \nu_e \bar{\nu} \nu$ and
$K^+ \rightarrow l^+ \nu_l \bar{\nu} \nu ~(l=e,\mu)$,
could modify the lepton energy spectra
in $K^+$ and $\pi^+$ decays. From an analysis of these
spectra the following bounds on the coupling $F$ were obtained~\cite{BBP70}
\begin{equation}
|F| \le 10^7 G_F,~~~~~~|F| \le 2\times 10^6 G_F~,
\label{firstbounds}
\end{equation}
where $G_F$ denotes the weak Fermi constant.
\par

Similar bounds were found \cite{BBP70} from the absence of leptons
with ``wrong'' charge in the process
$\nu_{\mu}~+~N \rightarrow \mu^+~+~ \nu_\mu~+~ \nu_\mu~+X$.
\par
Later on these bounds were improved in a special experiment \cite{CHP}
searching for the decay
$K^+ \rightarrow \mu^+ \nu_{\mu} \bar{\nu} \nu $.
 From the negative result of this experiment the following limit was set:
\begin{equation}
F \le 1.7\times 10^5 G_F~.
\label{nextbounds}
\end{equation}
\par
The reason why bounds on the non-standard neutrino interaction coming
from low-energy experiments are so loose is evident.
The SNI contributes only to the decays
with four particles in the final state, and such processes
are strongly suppressed by phase space compared with the standard
leptonic $\pi$ and $K$ decays.
\par
In ref. \cite{BBS93} the width of the decay
$Z \rightarrow \nu \bar{\nu} \nu \bar{\nu}$
was calculated in the presence of a non-standard $\nu$--$\nu$
interaction of the general vector and axial-vector form:
\begin{equation}
{\cal L}^{\nu-\nu} = F \sum_{i,j=e,\mu,\tau}
(\bar{\nu}_{i} O_{i\alpha} \nu_{i})
(\bar{\nu}_{j} O^\alpha_{j} \nu_{j})\ ,
\label{inter}
\end{equation}
where
\begin{equation}
O_i^\alpha= a_i \gamma^\alpha P_L
+ b_i \gamma^\alpha P_R\ ,
\label{oi}
\end{equation}
$P_L=\half(1-\gamma_5)$ and $P_R=\half(1+\gamma_5)$
are the left and right chirality projectors and
$F,a_i,b_i$ are real parameters. The coupling $F$ has dimension
$[M]^{-2}$.
 From the invisible width of the $Z$-boson, which has been
precisely  measured at LEP, and assuming three generations
of light neutrinos and lepton universality, the following bounds
were obtained \cite{BBS93} ($\tilde{F}=Fa_l^2$):
\begin{equation}
|\tilde{F}| \le 390 G_F~,
\label{ourbound1}
\end{equation}
in the case of the $V-A$ structure of the non-standard interaction and
\begin{equation}
|\tilde{F}| \le 710 G_F~
\label{ourbound2}
\end{equation}
for the pure vector case.
As the phase-space suppression is much smaller in the case of
the $Z$-boson decay, these last bounds are much better than those
obtained from $\pi$- and $K$-meson decays.

Taking all previous bounds, however, the ``secret'' effective
four-neutrino interaction could still be much stronger than the one
predicted by the minimal SM\footnote{Some
information about $\nu$--$\nu$ interactions
was obtained also from astrophysical
data\cite{KT87,Man87,Pal89}.
The corresponding bounds are weaker than those of \eq{ourbound1}
and \eq{ourbound2}.  Bounds coming from
primordial nucleosynthesis can be much stronger \cite{MT94} when the
four-neutrino interaction involves both left- and right-handed neutrinos. },
$\tilde{F}=G_F/\sqrt{2}$.
All previous bounds are
extracted from processes in which the new interaction is the only
relevant one and, therefore, observables depend quadratically
on $\tilde{F}$.
Obviously if the new interaction enters in loop corrections
to a SM process,  modifications come through its interference
with the SM amplitude and, then, the deviations from the SM predictions
will depend linearly on the coupling $\tilde{F}$.

For example, $\nu$--$\nu$ interactions will
contribute to the decay $Z \rightarrow \bar{\nu} \nu$ at the one-loop level
(see Figs. 1a and b)
and consequently to the invisible width of the $Z$-boson.
It is very simple to estimate the order of magnitude of
the corresponding one-loop corrections:
\begin{equation}
\frac{\Delta \Gamma_{\bar{\nu} \nu}}{\Gamma_{\bar{\nu} \nu}} \approx
\frac {\tilde{F} M_Z^2}{(4\pi)^2}~.
\end{equation}
As the invisible width of the $Z$-boson is now measured with an accuracy
better than 1\% \cite{LEP}, one finds the following bound
on the non-standard coupling $\tilde{F}$:
\begin{equation}
\tilde{F} \le (1\mrm{--}10)\cdot G_F.
\label{estimate}
\end{equation}
Clearly, from this rough estimate and from our previous discussion
one expects stronger bounds
on the ``secret'' $\nu$--$\nu$ interaction coming from
the one-loop analysis than those which follow from its contribution to
the invisible width of the $Z$-boson
at tree-level, \eq{ourbound1} and \eq{ourbound2}.
\par
The above estimate of the loop effects
of the four-neutrino interactions is rather na\"{\i}ve, because the
one-loop calculation is actually divergent: four-neutrino interactions are
not renormalizable. This does not preclude us obtaining some
information on them from loops, as long as the appropriate framework is
used to obtain finite non-ambiguous results. This framework is the
effective quantum field theory (EQFT) \cite{Wei79,EQFT}.
In this language all operators
allowed by the symmetries of the problem
are already present in the effective Lagrangian
from the beginning, therefore, there always exists a counterterm available to
absorb any divergence that could appear in loop
calculations. The number of the effective operators is generally infinite and
experimental observables depend on an infinite number of unknown
couplings\footnote{Note, however, that if the effective theory is a
low-energy limit of some known  renormalizable theory (see e.g. \cite{BS93})
all effective couplings can be expressed in terms of the few parameters of
the underlying theory.}.
However, the effects of higher-dimension operators are
suppressed in low-energy processes and one can truncate the Lagrangian
by keeping only a finite number of operators. Using the EQFT language
one has a well-defined prescription to calculate loop effects of
non-renormalizable operators. The price that has to be paid is that
it is not possible to analyse effects of one operator independently
of  other operators that mix with it under renormalization.
Under certain assumptions one can reduce the basis of operators
that mix.The more assumptions one
makes the stronger will be the bounds one
obtains on the couplings of the effective operators.
The less assumptions one
makes the more reliable will be the bounds obtained.

For example, if we want to analyse an operator that contributes to
experimental observables at the one-loop level we can use a ``minimal''
set of operators (which, in general, does not form a closed basis)
that contains the operator in question plus all the operators that
mix ``directly'' with it at the one-loop level\cite{BS93}.

In this letter we discuss the procedure of bounding
effective operators by considering the
case of the four-neutrino contact interaction and obtain constraints on
this elusive interaction from the processes in which it contributes at the
one-loop level.

\section{Effective $Z\bar{\nu}\nu$ vertex in the presence of the
four-neutrino interaction}
\label{effectivevertex}

We first calculate one-loop corrections
to the $Z$-neutrino coupling
due to a SNI of the general $V,A$
form given in \eq{inter}.
Note that interactions mediated by scalars can
also be written in this form after a Fierz transformation. The flavour
structure could be, however, more general. For simplicity we will only
consider the flavour structure of \eq{inter}. On the other hand,
it has been shown very recently \cite{MT94} that four-neutrino
interactions that involve both,  left-handed and right-handed
neutrinos are strongly bounded by cosmological arguments: if
these kind of interactions are strong enough they would keep the
three right-handed neutrinos in thermal equilibrium at the time of
nucleosynthesis, therefore, disturbing it. But it is important to realise,
that nucleosynthesis gives  no bound at
all if either $a_i$ or $b_i$ are zero in \eq{oi},
because in both cases right-handed neutrinos are completely
decoupled.  Interactions of only right-handed neutrinos are not
interesting, therefore, in the analysis
we could restrict ourselves to interactions
among only left-handed neutrinos, that is $a_i\not=0$ and $b_i=0$.
However, for the sake of generality we will keep for the moment
the two interactions as given by \eq{oi}.

Finally we would like to remark that
interactions of the form \rfn{inter} are not $SU(2)$ gauge invariant
by themselves. To write them in an explicitly gauge invariant form we would
need to include an additional interaction among charged leptons
(and neutrino--leptons as well)
with exactly the same coupling. But such interactions,
at least those involving electrons,  are strongly bounded by
different experiments. Therefore, if four-neutrino interactions
are part of an $SU(2)$ gauge invariant interactions
they can be bounded indirectly though
the bounds on the four-fermion interactions involving charged
leptons.  On the other hand, the general effective Lagrangian, \eq{inter},
can be obtained from a gauge-invariant effective Lagrangian
after spontaneous symmetry breaking.
 From a phenomenological point of view, however, we
could ask how strong four-neutrino interactions can be
independently of any additional assumption.
\par
We will use the na\"{\i}ve dimensional regularization scheme
(with anticommuting $\gamma_5$). Then, there are two one-loop
diagrams, shown in Figs. 1a and b, contributing to the
$Z\bar{\nu}\nu$ vertex\footnote{There
is no wave-function renormalization of the external neutrinos in
our case. Such massless tadpole-like diagrams are zero in dimensional
regularization.}.
The corresponding amplitudes are given by
\begin{eqnarray}
&& T_a(Z\rightarrow \nu_i \bar{\nu}_i)=
-\frac{g}{2c_W}\mu^\epsilon \frac{F}{(4\pi)^2} \frac{4}{3}
q^2\left(\Delta_\epsilon(q^2)+\frac{5}{3}\right)
\sum_{j=e,\mu,\tau} a_{j}
\bar{u}(k') O^\alpha_i u(k) \epsilon(q)_\alpha~, \\
&& T_b(Z\rightarrow \nu_i \bar{\nu}_i)=
-\frac{g}{2c_W}\mu^\epsilon \frac{F}{(4\pi)^2} \frac{4}{3}
q^2\left(\Delta_\epsilon(q^2)+\frac{2}{3}\right)
a_i^2\bar{u}(k') \gamma^\alpha P_L u(k) \epsilon(q)_\alpha~.
\label{fig}
\end{eqnarray}
Here $q=k'-k$ is the four-momentum of the $Z$-boson,
$c_W=\cos\theta_W$ is the cosine of the weak mixing angle,
and the summation in $T_a$ runs over the
different neutrino-types in the loop; $u(k)$ denotes a Dirac spinor,
$\epsilon_\alpha$ is the wave function of the $Z$-boson.
Finally, $\mu$ is the dimensional regularization mass parameter
and $\epsilon = 2-D/2$ with $D$ the space-time dimension.
Both diagrams are divergent.
In dimensional regularization this divergence
appears as a simple pole, $1/\epsilon$, in the function
\begin{equation}
\Delta_\epsilon=\frac{1}{\epsilon}-\gamma+\log(4\pi)
-\log\left(\frac {-q^2-i\eta}{\mu^2} \right)
\equiv\frac{1}{\hat{\epsilon}}
-\log\left(\frac {-q^2-i\eta}{\mu^2} \right).
\end{equation}
\par
\par
 In our analysis we will assume lepton universality for the
three generations of neutrinos. Then, we can rewrite the sum
of $T_a$ and $T_b$ in the form
\begin{equation}
T =T_a+T_b=-\frac{g}{2c_W}\mu^\epsilon G_F q^2
\sum_{A=L,R}
c_1^{A}(\gamma_{12}^{A} \Delta_\epsilon(q^2)+\kappa_{12}^A)
\bar{u}(k') \gamma^\alpha P_A u(k) \epsilon(q)_\alpha~,
\label{amplitude}
\end{equation}
which has been expressed in terms of the renormalized (scale-dependent)
dimensionless
couplings
\begin{equation}
c_1^L=\frac{Fa^2_i}{G_F},~~~~~~~~
c_1^R=\frac{Fa_i b_i}{G_F}
\end{equation}
and the following constants
\begin{equation}
\gamma_{12}^L=\frac{1}{3\pi^2}, ~~~
\gamma_{12}^R=\frac{1}{4\pi^2}, ~~~
\kappa_{12}^L=\gamma_{12}^L\frac{17}{12}, ~~~
\kappa_{12}^R=\gamma_{12}^R\frac{5}{3}~ .
\end{equation}
\par
As we see, the SNI
generates, at the one-loop level, a derivative coupling of the $Z$-boson
to neutrinos that is divergent. In order to obtain a finite amplitude
for processes involving the $Z\bar{\nu}\nu$ vertex,
the effective Lagrangian should contain a
term able to absorb this divergence in its coupling. This term can
be written as
\begin{equation}
{\cal L}_2=-\frac{g}{2c_W} \mu^\epsilon G_F
\sum_{i=e,\mu,\tau}\sum_{A=L,R}
\left(c_2^{A}+\Delta c_2^A\right)
(\bar{\nu}_{i} \gamma^\alpha P_{A}\nu_{i} )
\partial^\beta Z_{\beta\alpha} ~,
\label{ct}
\end{equation}
where
$Z_{\beta\alpha}=\partial_\beta Z_\alpha -\partial_\alpha Z_\beta$;
$c_2^A(\mu)~(A=L,R)$ are the $\overline{\mrm{MS}}$
renormalized couplings and the corresponding counterterms
are
\[
\Delta c_2^A = - c_1^A \gamma_{12}^A \frac{1}{\hat{\epsilon}}~.
\]


It is important to remark that since we are not assuming
$SU(2)$ invariance for the effective Lagrangian,
 a direct (non-derivative) non-standard coupling of the
$Z$ boson to neutrinos is in principle possible. Such a coupling could
be generated from a gauge-invariant effective Lagrangian after spontaneous
symmetry breaking. Although in dimensional
regularization and for massless neutrinos it is not needed,  there is no
symmetry that forbids it. In fact it will appear naturally if other
regularization schemes are used. Therefore, in order to be completely
general we will include it in the analysis and will see how
it affects our bounds. This
additional interaction has the form
\begin{equation}
{\cal L}_\delta=-\frac{g}{2c_W} \mu^\epsilon
\sum_{i=e,\mu,\tau}\sum_{A=L,R}\delta^A Z_\alpha
(\bar{\nu}_{i} \gamma^\alpha P_{A}\nu_{i} )~.
\label{delta}
\end{equation}
The contribution of the operators \rfn{ct} and \rfn{delta}
is schematically shown in Fig.~1c.
Then, the full renormalized $Z\bar{\nu}\nu$ vertex will be given
by the sum of the three diagrams of Fig.~1:
\par
\begin{equation}
\label{renamp}
\hat{T} =-\frac{g}{2c_W}  \sum_{A=L,R} g_{A}(q^2)
\bar{u}(k') \gamma^\alpha P_A u(k) \epsilon(q)_\alpha~,
\end{equation}
where
\begin{equation}
g_A(q^2)=\delta^A(\mu)+G_F q^2 \left(c_2^A(\mu)+
c_1^A(\mu) \left(\gamma^A_{12}
\left(\log(\mu^2/|q^2|)+i\pi\theta(q^2)\right)
+ \kappa_{12}^A\right)\right)~,~~~A=L,R~.
\label{deltag}
\end{equation}
Here, $\delta^A(\mu)$  gives the contribution of the direct non-derivative
$Z$-neutrino interactions. The running couplings in our approximation
(we neglect all contributions with gauge bosons running in the loops) are
given by
\begin{eqnarray}
&& c_1^A(\mu) \approx c_1^A(\mu_0) \\
&& c_2^A(\mu) \approx c_2^A(\mu_0) + c_1^A(\mu_0) \gamma^A_{12}
\log \left( \frac{\mu_0^2}{\mu^2}\right)~,
\end{eqnarray}
where $\mu_0$ is some reference scale.
Thus, the effective four-neutrino operator
at the one-loop level contributes to the running of the coupling
of the operator \rfn{ct} and we have to consider mixing between at least
these two operators\footnote{Obviously, there are
many other four-fermion operators like
$(\bar{l}l)(\bar{\nu}\nu)$, etc., which also mix with the $Z$-neutrino
coupling \rfn{ct}. But as we are neglecting loops with gauge bosons,
they do not mix directly at the one-loop level with
the four-neutrino operator and, as they can be strongly bounded
from other processes, we will disregard them.}.
The coupling $\delta(\mu)$
does not mix with the other couplings because it correspond to
an operator of different dimension, then
$\delta^A(\mu) \approx \delta^A(\mu_0)$.

On the other hand, because the standard $Z\bar{\nu}\nu$ coupling
only involves left-handed neutrinos,
the lowest-order effect of the non-standard vertex \eq{renamp} comes
via its interference with the standard coupling and therefore only
the real part of the left-handed vertex in \eq{renamp} will be relevant
for our analysis. Thus, ``new physics'' depends on three unknown parameters,
$\delta^L(\mu)$, $c_1^L(\mu)$ and $c_2^L(\mu)$.
In addition, as commented above, couplings to
right-handed neutrinos are strongly bounded from cosmological data.

Obviously, to put independent bounds on these
three parameters in processes
which depend only on the induced effective $Z$-neutrino vertex
one needs experimental information obtained at least at three
different energy scales. We would like to note that the behaviour of the
three terms is quite different, while the terms proportional
to $c_1^L$ and $c_2^L$ are also proportional to $q^2$ the $\delta^L$
term is independent of $q^2$ and can be bounded at very
low energies.

\section{Neutrino neutral-current experiments and \newline
 bounds on the four-neutrino contact interaction}

In this section we will consider the bounds on the four-neutrino
coupling (and other effective coupligns entering in the
$Z\nu\nu$ vertex \rfn{renamp}),
which follow from neutrino neutral-current experiments.
\par
As we discussed above, in our approximation, the effect
of the non-standard
operators is taken into account by three renormalized coupling constants
$\delta^L(\mu)$, $c_1^L(\mu)$ and $c_2^L(\mu)$.
Since physical results are independent from $\mu$ we can freely choose this
scale. We will take $\mu=M_Z$ as the reference scale and define, for further
use,
\begin{equation}
\delta=\delta^L(M_Z)~,~~~~c_1=c_1^L(M_Z)~,~~~~c_2=c_2^L(M_Z)~,~~~~
\gamma=\gamma_{12}^L=\frac{1}{3\pi^2}~,~~~~
\kappa=\kappa_{12}^L = \frac{17}{36\pi^2}~.
\end{equation}
\par
Then all our observables will depend on the quantity
\begin{equation}
\Re{g_L(q^2)} = \delta+G_F q^2 \left(
c_2+c_1 \kappa+c_1 \gamma \log(M_Z^2/|q^2|)
\right)~.
\label{regl}
\end{equation}

We will first consider bounds from the precise measurement of the
invisible $Z$-width at LEP.
 From the effective vertex \rfn{renamp}
we can easily obtain the partial decay width of the $Z$-boson
into two neutrinos. It can be written in the following form:
\begin{equation}
\Gamma(Z\rightarrow \bar{\nu} \nu) =
\Gamma^{SM}(Z\rightarrow \bar{\nu} \nu)
+\Delta \Gamma_{\bar{\nu} \nu}~,
\label{gamnunu}
\end{equation}
where $\Gamma^{SM}(Z\rightarrow \bar{\nu} \nu)$
is the SM contribution (including radiative corrections)
and $\Delta \Gamma_{\bar{\nu}\nu}$ contains the effects of the non-standard
operators.
At lowest order it comes from the interference of the
non-standard amplitude with the SM amplitude and we have
\begin{equation}
\Delta \Gamma_{\bar{\nu} \nu}~=~
\Gamma^{SM}(Z\rightarrow \bar{\nu} \nu) 2 \Re{g_L(M_Z^2)},
\label{dgamnunu}
\end{equation}
where the function $\Re{g_L(q^2)}$ is
given by \eq{regl} and for $q^2=M_Z^2$ it is
\begin{equation}
\Re{g_L(M_Z^2)}=\delta+G_F M_Z^2 (c_2+c_1 \kappa)~.
\end{equation}
\par
Assuming that there are three generations of neutrinos, the non-standard
contribution to the invisible width of the $Z$-boson is
\begin{equation}
\Delta \Gamma_{invis}=
3\Delta \Gamma_{\bar{\nu} \nu}~,
\label{obs1}
\end{equation}
where $\Delta \Gamma_{\bar{\nu} \nu}$ is given by \eq{dgamnunu}.
On the other hand this quantity can also be expressed as
\begin{equation}
\Delta \Gamma_{invis}=
\Gamma_{invis} -
3 \left(\frac{\Gamma_{\bar{\nu} \nu}}{\Gamma_{\bar{l} l}}\right)^{SM}
\Gamma_{\bar{l} l}~.
\label{dginv}
\end{equation}
The r.h.s. of \eq{dginv} is constructed only from observables measured
at LEP \cite{LEP}:
\begin{equation}
\Gamma_{invis} = 497.6 \pm 4.3~\mrm{MeV},~~~~~
\Gamma_{\bar{l} l}=83.87 \pm 0.27~\mrm{MeV}
\label{lepdata}
\end{equation}
(we use the combined result from the four LEP experiments \cite{LEP})
and the ratio of the neutrino and charged leptons partial widths
calculated within the SM
\begin{equation}
\left(\frac{\Gamma_{\bar{\nu} \nu}}{\Gamma_{\bar{l} l}}\right)^{SM}
= 1.992 \pm 0.003~.
\label{smratio}
\end{equation}
The central value of the above quantity corresponds to $m_{top}=150$~GeV
and the small error is due to the variation
of the mass of the top quark in the range
$100~\mrm{GeV} < m_{top} <200~\mrm{GeV}$.
Using \rfn{lepdata} and \rfn{smratio} we obtain
\begin{equation}
\Delta \Gamma_{invis} \simeq -4 \pm 5~\mrm{MeV}~.
\label{deltagam}
\end{equation}
 From this experimental result and from
our calculation of the extra contributions to the invisible $Z$ decay
width we obtain
\begin{equation}
\label{nnbounds}
-0.009 \le \delta+G_F M_Z^2(c_2 +c_1 \kappa) \le 0.001~.
\end{equation}
It is obvious that from this
equation we cannot get bounds on all three couplings $\delta$,
$c_1$ and $c_2$ unless additional assumptions are considered.
In equation~\rfn{nnbounds} one can consider two situations:
\begin{enumerate}
\item
There are no unnatural cancellations among the three
terms $\delta$,  $\kappa G_F M_Z^2 c_1$ and
$G_F M_Z^2 c_2$. Then each of them can be
bounded independently of the others and we obtain:
\begin{equation}
\abs{c_1} \le \frac{0.009}{\kappa G_F M_Z^2 } = 2,\bla
\abs{c_2} \le \frac{0.009}{G_F M_Z^2} = 0.09,\bla
\abs{\delta} \le 0.009
\label{natbounds}
\end{equation}
\item
$\delta \approx G_F M_Z^2 c_2 \approx  \kappa G_F M_Z^2 c_1$.
In this case
there could be cancellations among the three terms.
However, even
if there are cancellations
at this particular scale ($M_Z$) there will be no cancellations at
other scales. In what
follows we will show that, also in the case of cancellations at LEP
energies, it is still possible to get some interesting
bounds on the coupling $c_1$ if additional data obtained at different
scales are used.
\end{enumerate}

Several types of experiments are sensitive to the neutral current
neutrino interactions at different $q^2$ scales.
As the non-standard operators \rfn{inter} and \rfn{ct}
contribute to the derivative
$Z\bar{\nu}\nu$ coupling and this
contribution is proportional to $q^2$ we can get some reasonable
additional information
on the couplings $c_1$ and $c_2$ only from DIS experiments
at high energy $(-q^2\simeq 100-1000~\mrm{GeV}^2)$.
However,  the direct $Z$-neutrino non-derivative interaction,
given by $\delta$,  contributes with the same strength to any energy,
therefore, it can be  bounded also in low-energy experiments,
e.g. in the elastic $\bar{\nu},~\nu$--electron scattering
$(-q^2\simeq 10^{-2}~\mrm{GeV}^2)$.
Since DIS experiments
are more precise and are performed at different energy scales,
we will mainly use their results in our analysis.

There are several high-precision DIS measurements
\cite{CDHS,CHARM,CCFR}, CDHS, CHARM and CCFR. The first two
experiments are performed with neutrino beam peek energies of
about  50 GeV and the same energy spectrum,
while CCFR operates with an average energy of  161 GeV.
As we will see this gap
in energies will be enough four our purposes.

The most interesting quantity measured in DIS experiments
is the ratio of the neutral-current to charged-current cross sections
for neutrino  beams
\begin{equation}
R_\nu=\frac{\sigma^{NC}_\nu}{\sigma^{CC}_\nu}~.
\label{rrat}
\end{equation}
Again as in a case of $\Gamma(Z\rightarrow \bar{\nu}\nu)$
the theoretical prediction for this quantity can be written
as a sum of the standard and the non-standard contributions:
\begin{equation}
R_\nu=R^{SM}_\nu+\Delta R_\nu~.
\end{equation}
Using our effective  $Z \bar{\nu} \nu$ vertex, \eq{renamp}, we obtain
\begin{equation}
\Delta R_\nu =
\frac{2\int dx \int dy \frac{d\sigma^{NC}_\nu}{dx dy} \Re{g_L(q^2)} }
{\int dx \int dy \frac{d\sigma^{CC}_\nu}{dx dy}}~,
\label{deltar}
\end{equation}
where $d\sigma^{NC}_\nu/dxdy$ and $d\sigma^{CC}_\nu/dxdy$
are differential neutral- and charged-current cross sections
calculated within the SM, and $x$ and $y$ are usual DIS variables.
The momentum transfer squared is given by
\begin{equation}
q^2=-2 E_{beam} M_p x y~
\end{equation}
with $E_{beam}$ the beam energy in the laboratory frame and $M_p$
the proton mass. The function $\Re{g_L(q^2)}$ is defined in \eq{regl}.
\par
The final result of the analysis of the experimentally measured
ratios $R_\nu$
is usually presented in terms of the
value of weak mixing angle $\sin^2\theta_W=1-M_W^2/M_Z^2$. The
quoted values are\footnote{The actual values we use are taken from a recent
publication of the CCFR collaboration \cite{CCFR};
there, $\sin^2\theta_W=1-M_W^2/M_Z^2$ is given for the
same masses of the charm quark,
$m_c=1.31\pm 0.24~\mrm{GeV}$, and the top quark $m_t=150~\mrm{GeV}$
for all experiments\cite{CDHS,CHARM,CCFR}.}:
\begin{eqnarray}
\mrm{CDHS}~~\mrm{\cite{CDHS}}~&:&~~0.2225 \pm 0.0066 \nonumber\\
\mrm{CHARM}~\mrm{\cite{CHARM}}~&:&~~0.2319 \pm 0.0065 \nonumber\\
\mrm{CCFR}~~\mrm{\cite{CCFR}}~&:&~~0.2218 \pm 0.0059~.
\label{dissin}
\end{eqnarray}
On the other hand the same quantity can be obtained from
LEP and collider (UA2, CDF) data with very high
precision. The average is \cite{LEP}
\begin{equation}
\mrm{LEP} + M_W~:~~\sin^2 \theta_W^{LEP} = 0.2255 \pm 0.0005~.
\label{lepsin}
\end{equation}
\par
Using $\sin^2 \theta_W=1-M_W^2/M_Z^2$ as an input, one can
obtain the predictions for $R_{\nu}$,
which in the case of neutrinos scattered off an approximately isoscalar
target are given by the tree level expression \cite{LLE83}
\begin{equation}
R_\nu^{SM}=\frac{1}{2}-s_W^2+\frac{5}{9}s_W^4\left(1+R^{cc}\right)~,
\label{lle}
\end{equation}
where
$R^{cc}=\sigma^{CC}_{\bar{\nu}}/\sigma^{CC}_{\nu} \approx 0.4$
is the ratio of the antineutrino--neutrino charged-current cross sections.
\par
In our analysis we neglect radiative corrections and parton-model
corrections to the non-standard contribution  $\Delta R_\nu$.
Then, the deviations from the standard result
are given by the
difference between the values of $R_\nu$ obtained by using
the $\sin^2 \theta^i_W$ measured in  DIS experiments and those
obtained by using  $\sin^2 \theta^{LEP}_W$ measured at LEP.
For every experiment we have:
\begin{equation}
\Delta R_{\nu}^i=
R_{\nu}(\sin^2 \theta_W^i)- R_{\nu}(\sin^2 \theta_W^{LEP})~,~
\label{drexp}
\end{equation}
where $\Delta R_{\nu}^i$ is given by \eq{deltar} and the index
$i=${\small CDHS, CHARM, CCFR}
refers to the different conditions of the different experiments (beam energy
and the cut on the $y$-variable), which influence the calculation \rfn{deltar}.
In \eq{drexp} the predictions $R_\nu(\sin^2 \theta_W)$ are
calculated according to the tree-level expression \rfn{lle} and using
values for $\sin^2\theta_W$ given by \rfn{dissin}
and \rfn{lepsin}. Then, for $\Delta R_{\nu}^i$ we have:
\begin{eqnarray}
\mrm{CDHS}~~&:&~+0.0022 \pm 0.0046 \\
\mrm{CHARM}~&:&~-0.0039 \pm 0.0044 \\
\mrm{CCFR}~~&:&~+0.0026 \pm 0.0042 \label{ccfr}~.
\label{rdis}
\end{eqnarray}

Using these experimental values we can obtain the bounds on the
coupling $c_1$, even if there are cancellations among the different
couplings. Before doing the complete numerical analysis we will do
a simple estimate of the bounds, which, as we will see, works very well.
For this estimate we rewrite \eq{deltar} as
\begin{equation}
\Delta R_\nu = R_\nu 2\Re{g_L(q_i^2)}~,
\label{deltara}
\end{equation}
where $q_i^2$ is is some effective average of $q^2$ for the experiment
chosen in order to reproduce the complete result.
We use that CDHS and CHARM experiments
are performed  with the same neutrino beam, then we average their results
and obtain
\begin{equation}
CERN=CDHS+CHARM:~~~\Delta R_\nu = -0.00085 \pm 0.0032~,~~~~~~~
\abs{q^2_{N}} \approx 14~\mrm{GeV}^2~.
\label{vcern}
\end{equation}
For CCFR we have
\begin{equation}
CCFR:~~~\Delta R_\nu = +0.0026 \pm 0.0042~,~~~~~~~
\abs{q^2_{R}} \approx 45~\mrm{GeV}^2~.
\label{vccfr}
\end{equation}
Then, the bounds from the different experiments can be expressed as
\begin{eqnarray}
&\abs{\delta+G_F M_Z^2 (c_2+c_1 \kappa) }  & \le  b_{L}\label{blep}\\
&\abs{\delta+G_F q_{R}^2 (c_2+c_1 \kappa+c_1 \gamma
\log(M_Z^2/|q_{R}^2|))}  & \le   b_{R}\label{bccfr}\\
&\abs{\delta+G_F q_{N}^2 (c_2+c_1 \kappa+c_1 \gamma
\log(M_Z^2/|q_{N}^2|))} & \le  b_{N} \label{bcern}~,
\end{eqnarray}
 From the estimates in \eq{nnbounds}, \eq{vcern} and \eq{vccfr}, and
taking into account that $R_\nu \approx 0.314$ we have
\begin{equation}
b_{L}=0.009~,~~~~~~~b_{R}=0.0110~,~~~~~~~b_{N}=0.0051~,
\end{equation}

Using error propagation, form eqs.~\rfn{blep}--\rfn{bcern} we can extract
bounds  for the different couplings $\delta$, $c_2$ and  $c_1$.
Taking into account that  $\abs{q^2_{N}} , \abs{q^2_{R}} \ll M_Z^2$ we
obtain
\begin{eqnarray}
\abs{\delta} & \le & \left(b_{N}/(1-z)\right)\sqrt{1-z^2 b_{R}/b_{N}}=0.012
\label{bound1}\\
\abs{c_2} & \le & \left.\kappa\sqrt{b_{N}^2+b_{R}^2}\right/
\left(\gamma G_F |q_{R}^2| \log(M_Z^2/|q_{R}^2|) (1-z)\right) = 10
\label{bound2}\\
\abs{c_1} & \le &\left.\sqrt{b_{N}^2+b_{R}^2} \right/
\left(\gamma G_F |q_{R}^2| \log(M_Z^2/|q_{R}^2|) (1-z)\right) =222~.
\label{bound3}
\end{eqnarray}
Here $z= \left.\left(|q_N^2|\log(M_Z^2/|q_N^2|)\right)\right/
\left(|q_R^2|\log(M_Z^2/|q_R^2|)\right)
\approx 0.4$. These are
reliable bounds in the case of cancellations among the
contributions of the different operators.
\par
In the complete analysis, we did a three-parameter fit, in $\delta,c_2,c_1$,
of the theoretical expressions \eq{dgamnunu} and \eq{deltar} to the
data.
We would like to note that in the numerical calculation we used rather
simple parametrizations of the parton distribution functions \cite{DO}.
However,  different choices of the structure functions
do not change our results noticeably. The reason for this is that our
non-standard contributions are mainly sensitive to the
parton distributions
at large values of the $x$ and $y$ variables.
\par
As a result of the three-parameter fit to the full body of data
we obtain the following bounds at $68\%$~C.L.
\begin{eqnarray}
\delta =& 0.004 \pm 0.009~~\Longrightarrow &    \abs{\delta}~ \ltap~ 0.013\\
c_2=& 4.7 \pm 7~~~~~\Longrightarrow &  \abs{c_2} \ltap~ 12\\
c_1=& -100 \pm 140~~~~\Longrightarrow
& \abs{c_1} \ltap~ 240 \label{c1bound}~.
\end{eqnarray}
These constraints are in a good agreement with our estimate in
eqs.~\rfn{bound1}--\rfn{bound3}.
The extreme values of $\tilde{F}$, of order $\sim 240 G_F$, are possible
only because of large
cancellations between the contributions of the
three non-standard couplings.
If one decides that such cancellations are unnatural,
then one obtains a much better bound for the contact four-neutrino
interaction. The complete analysis gives in this case
\begin{equation}
           \abs{\tilde{F}} \ltap~ 2 G_F~.
\end{equation}
\par
The above bounds can be improved in the future. In the
case of cancellation between the different couplings,
the bounds are defined
essentially by the errors of the experiments at lower energies; therefore,
only better DIS data can improve the bound in \eq{c1bound},
especially if DIS experiments are performed at higher energies.
If there are no cancellations between the different couplings, the higher scale
experiment is the relevant one, and future improvements
of the measurement
of the invisible width of the $Z$ at LEP will be very important.
\section{Conclusions}

In this letter we have obtained new constraints on non-standard
four-neutrino interactions coming from their contribution
at the one-loop level to the
invisible width of the $Z$-boson and to deep inelastic scattering.
The bounds obtained from a conservative model-independent analysis
of the ``secret'' neutrino interactions at the one-loop level
improve at least by a factor 2 previous constraints,
\rfn{ourbound1} and \rfn{ourbound2}, that were
obtained from the study of the non-standard
$Z\rightarrow 4\nu$ decay. If
there are no unnatural cancellation between the contributions of the
various non-standard couplings,
a much stronger bound on the strenght of a four-neutrino interaction
has been obtained. This bound is 200 times better than previous
constraints.

\section*{Acknowledgements}
It is a pleasure for us to thank F. Botella,
S. Peris and H. Spiesberger for helpful discussions and
A. Bodek for providing us with updated CCFR data. One of us (M.B.)
is indebted to the CERN TH Division for its kind hospitality.

\section*{Figure captions}
{\bf Figure 1a--c:} Diagrams that give contributions
Diagrams that give contributions
to the $Z\bar{\nu} \nu$ vertex in the presence of the non-standard
four-neutrino interaction. In diagram (a), neutrinos of different
flavours are running in the loop.

\end{document}